# Least-square based recursive optimization for distance-based source localization

Thien-Minh Nguyen, Lihua Xie

*Abstract*— In this paper we study the problem of driving an agent to an unknown source whose location is estimated in real-time by a recursive optimization algorithm. The optimization criterion is subject to a least-square cost function constructed from the distance measurements to the target combined with the agent's self-odometry. In this work, two important issues concerning real world application are directly addressed, which is a discrete-time recursive algorithm for concurrent control and estimation, and consideration for input saturation. It is proven that with proper choices of the system's parameters, stability of all system states, including on-board estimator variables and the agent-target relative position can be achieved. The convergence of the agent's position to the target is also investigated via numerical simulation.

## I. INTRODUCTION

In research on control of multi-agent, an almost undeniable ingredient in the formulation of any control law is the neighbour's state, which usually is positioning information in swarm systems [1]. However, in most practical scenarios, this information is inaccessible due to cost, limited payload capacity, or communication capability. In fact, many works have been proposed in the literature to resolve this problem by estimating the neighbour's position [2]–[4]. Inspired by this issue, in this paper we propose a technique to estimate the position of a target using only distance measurements with potential application in many mobile swarm systems.

Target localization and tracking using distance measurements is a problem that has been studied extensively and several approaches have been put forth in the literature. In one of the earliest discussions on the problem, Dandach et al. [2] proposed a gradient-based adaptive estimation algorithm to estimate the location of source using distance measurements and the agent's positioning information. Later, a least-square-based algorithm was proposed in [3] to improve the performance. Using the same adaptive scheme, in [5]–[7] the authors substituted the requirement for knowledge on the agent's position with velocity and derivative of distance to estimate the relative position between agents and a single landmark. However, in these works asymptotic convergence of estimation error requires that the relative velocities are *persistently exciting*. Therefore the agent has to terminate the adaptive localization process after some time before approaching the target using other control strategies. Notable works with physical implementation using similar approach can be found in [8]–[10].

The authors are with School of Electrical and Electronic Engineering, Nanyang Technological University, Singapore 639798, 50 Nanyang Avenue.
The research is partially supported by the ST Engineering - NTU Corporate Lab through the NRF corporate lab@university scheme.
Email for correspondence: elhxie@e.ntu.edu.sg.

A more challenging approach, which is the one considered in this paper, is to drive the agent towards the target while estimating its location at the same time, as illustrated in Fig. 1. To the best of our knowledge, the works by [11]–[13] can be considered as pioneers of this approach. In [11], the control objective was to drive the agent to a circular trajectory whose radius around the target is a user-defined parameter. This objective makes sure that a certain level of persistent excitation still exists to help bound the tracking error when the target drifts. Some other works such as [14], [15] also use this control objective but substitute positioning information by velocity and range rate.

Different from the previous objective, an even more challenging control objective is tackled in [12]. In this work, the authors seek to achieve asymptotic convergence of the agent's position to the target's location. Their algorithm resolves the dichotomy between persisting excitation and waning control input when approaching the target. Using the same approach, *Güler et al.* in [13] generalized the estimation problem for the multiple-target case and modified the control law to reach a desired formation described by the distances to the beacons instead of explicit coordinates.

For the algorithm to be of practical use, there are two main concerns that we believe are important but haven't been addressed in previous works. First, an algorithm must be converted to or approximated by a discrete-time formulation for real implementation. So far, most of the algorithms in previous works are still formulated and analyzed in a continuous-time context. A discrete-time algorithm was given in [16], however no formal analysis was given. This observation motivates us to reformulate and directly analyze the problem in discrete time. From our analysis, several parametric conditions to guarantee the stability of the system are derived which are similar to the continuous-time case. Second, the existing works have not considered input saturation which may not be practically viable. In this work, the saturation effect is directly studied and plays an active role in guaranteeing the stability of the system.

A numerical example is used to verify the theoretical findings of the paper. Even when the target is no longer static as assumed, the distance to this target remains bounded. This suggests that the zero-value relative position is asymptotically stable and we are looking forward to definitely confirming this convergence in a future work. The theoretical proof of the convergence is left for future work.

The remainder of the paper is organized as follows: in Section II, we describe the design of the estimator and the control scheme in details. In Section III, we provide our anal-ysis on the stability and convergence of the system. Section IV presents our simulation to demonstrate the efficacy of the proposed algorithm. Finally, Section V concludes the paper and discusses some direction for future works.

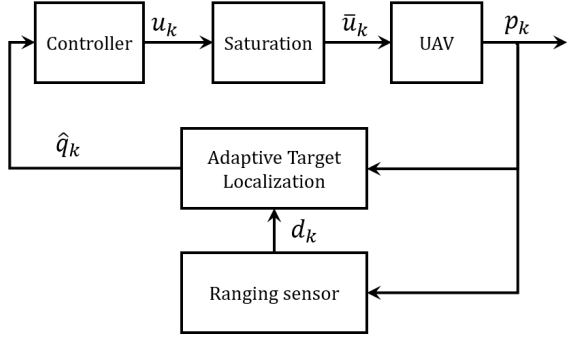

Fig. 1. Overview of adaptive target localization and tracking system.

## II. PROPOSED SYSTEM

Fig. 1 offers an overview of our system with unknown target's location as the input, and the output is the agent's position. Our goal is to drive the agent's position $p_k$ to the target's location $q$. The *Adaptive Target Localization* block plays the crucial role in the system by providing estimate of the target location $\hat{q}_k$ using the distance measurements and the agent's odometry information. Notice that we will assume that the target is static in formulating the algorithms.

### A. Dynamics model of the agent

Denote $\tau$ as the sampling period and $p_k$ as the agent's position at time step $k\tau$ and assume that the agent's dynamics satisfies the following discrete-time single integrator model:

$$p_{k+1} = p_k + \tau \bar{u}_k = p_k + \tau \eta(u_k, v_M), \quad (1)$$

where $v_M$ is the maximum distance the agent can travel within $\tau$, $\bar{u}_k$ is the so-called saturated control signal and $u_k$ is the so-called the non-saturated control signal. The saturation $\eta \colon \mathbb{R}^m \times \mathbb{R} \to \mathbb{R}^m$ is defined as follows:

$$\eta(a,b) = \begin{cases} a, & \|a\| \leq b, \\ b(a/\|a\|), & \text{otherwise}, \end{cases} \quad (2)$$

where $\|\cdot\|$ denotes the Euclidean norm.

### B. Estimator

Denote $q$ as the target's position. We have the following relationship between the target's position and the distance measurements $d_k$:

$$d_k^2 = (p_k - q)^\top (p_k - q) = \|p_k\|^2 + \|q\|^2 - 2q^\top p_k.$$

Hence,

$$d_k^2 - d_{k-1}^2 = \|p_k\|^2 - \|p_{k-1}\|^2 - 2q^\top (p_k - p_{k-1}). \quad (3)$$

Let us define the following signals:

$$\phi_k = p_k - p_{k-1}, \quad (4)$$

$$\zeta_k = \frac{1}{2}\left(\|p_k\|^2 - \|p_{k-1}\|^2 - d_k^2 + d_{k-1}^2\right). \quad (5)$$

From (3) and (1), one can verify that:

$$\phi_k = \tau \bar{u}_{k-1}, \quad (6)$$

$$\zeta_k = q^\top \phi_k. \quad (7)$$

Since $q$ is unknown, we are to design an estimator for it. Thus, we denote the state estimate of $q$ as $\hat{q}_k$ and the estimation error state as $\tilde{q}_k \triangleq \hat{q}_k - q$. For some initial guess $q_0 \in \mathbb{R}^n$ and some positive-definite $n$-dimensional matrix $\Gamma_0$, let us consider the following cost function

$$J_k = \frac{1}{2}\sum_{j=1}^{k}(\zeta_j - q^\top \phi_j)^2 + \frac{1}{2}(q-q_0)^\top \Gamma_0^{-1}(q-q_0). \quad (8)$$

In [17], a so-called *pure least-square* algorithm was proposed to estimate the parameter $q$ recursively when new observation, i.e, $\zeta_k$ and $\phi_k$, are obtained. One issue with the existing method is that the proposed *adaptive control gain* may become arbitrary small, thus, the convergence of the algorithm can be slowed down. Therefore, in this work we propose the following law with a *covariance resetting* operation to overcome such problem:

$$\varepsilon_k = \zeta_k - \phi_k^\top \hat{q}_{k-1}, \quad (9)$$

$$\Gamma'_k = \Gamma_{k-1} - \frac{\Gamma_{k-1}\phi_k \phi_k^\top \Gamma_{k-1}}{1 + \phi_k^\top \Gamma_{k-1}\phi_k}, \quad (10)$$

$$\Gamma_k = \begin{cases} \Gamma_0, & \lambda_{\min}(\Gamma'_k) < \gamma_2, \\ \Gamma'_k, & \text{otherwise}, \end{cases} \quad (11)$$

$$\hat{q}_k = \hat{q}_{k-1} + \Gamma_k \phi_k \varepsilon_k, \quad (12)$$

where $\Gamma_k \in \mathbb{R}^{n \times n}$, $0 < \Gamma_0 \leq \gamma_1 I$, $0 < \gamma_2 < \gamma_1$.

### C. Non-saturated control law

The design of our non-saturated control input $u_k$ is inspired by [12] as follows:

$$u_k = -\beta(p_k - \hat{q}_k) + f(d_k)\sigma_k, \quad (13)$$

where $\beta > 0$ and the terms $f(d_k)$ and $\sigma_k$ will be elaborated via some definitions.

**Definition II.1.** $f \colon \mathbb{R}^+ \to \mathbb{R}^+$ is a strictly increasing, bounded function satisfying $f(0) = 0$ and $f(x) < x, \forall x > 0$.

The idea and one instance of the function $f(x)$ was provided in [12]. Here we would like to introduce an extra argument $\lambda > 0$ to offer another degree of freedom for tuning,

$$f(d) = \frac{1}{\lambda}(1 - e^{-\lambda d}). \quad (14)$$

**Definition II.2.** $\sigma_k$ is a periodic process with period $N$ and spans $\mathbb{R}^n$ in its period. Moreover, $\|\sigma_k\| \leq \sigma_M$.

The design of the $\sigma_k$ term in the control law (13) is inspired by the persistent excitation condition which has been addressed in [12], [16], [3]. Intuitively, it is to have the agent's trajectory span all of the possible spacial directions so that we can avoid the case where estimation error $\tilde{q}_k$ converges to some value that is orthogonal with the UAV's trajectory. In this paper, $\sigma_k$ is chosen as follows:

$$\sigma_k = \begin{bmatrix} r_1 \cos(\omega k + \varphi) \\ r_1 \sin(\omega k + \varphi) \\ r_2 \cos(3\omega k + \theta) \end{bmatrix}, \quad (15)$$

for some $\varphi, \theta \in [0, 2\pi]$, $r_1, r_2 > 0$, $r_1^2 + r_2^2 < \sigma_M$ and $\omega = 2\pi/N$. Notice that if $N = 1, 2, 4, 6, 12$ then $\sigma_k$ may not be able to span $\mathbb{R}^n$ as required in Definition II.2. A specific value of $N$ is used in the simulation and has been analyzed to guarantee that $\sigma_k$ spans $\mathbb{R}^n$ for any value of $\varphi$, $\theta$.

## III. STABILITY

In this section, we provide our analysis on stability of the proposed system subject to the choice of the parameters.

### A. Stability of estimator states

**Theorem 1.** *If $\gamma_1(\tau v_M)^2 < 2$ then the update laws delineated in (9), (10), (11), (12) have the following properties:*

i. $\hat{q}_k, \tilde{q}_k \in \ell_\infty$, $\varepsilon_k \in \ell_\infty \cap \ell_2$,
ii. $\lim_{k \to \infty} \varepsilon_k = 0$,
iii. *if $\Gamma_0 = \gamma_1 I$ then $\|\tilde{q}_k\| \leq \|\tilde{q}_0\|, \forall k > 0$.*

*Proof.*

i. Notice that in (11) $\Gamma_k \leq \Gamma_{k-1}, \forall k \geq 1$. Thus, the *covariance resetting* operation used in (11) is to guarantee that $\gamma_2 I \leq \Gamma_k \leq \gamma_1 I$ and $\gamma_1^{-1} I \leq \Gamma_k^{-1} \leq \gamma_2^{-1} I, \forall k > 0$. By letting $A = \Gamma_k^{-1}$, $B = \phi_k$ and $C = \phi_k^\top$ in the following identity,

$$(A + BC)^{-1} = A^{-1} - A^{-1}B(I + CA^{-1}B)^{-1}CA^{-1} \quad (16)$$

the update law of $\Gamma_k$ through in (11) and (10) is equivalent with

$$\Gamma_k^{-1} = \begin{cases} \Gamma_0^{-1}, & \lambda_{\max}(\Gamma_{k-1}^{-1} + \phi_k \phi_k^\top) > \gamma_2^{-1} \\ \Gamma_{k-1}^{-1} + \phi_k \phi_k^\top, & \text{otherwise} \end{cases} \quad (17)$$

From (9) and (7), it can be seen that

$$\varepsilon_k = -\phi_k^\top \tilde{q}_{k-1} \quad (18)$$

Substituting this to (12) yields

$$\tilde{q}_k = (I - \Gamma_k \phi_k \phi_k^\top) \tilde{q}_{k-1} \quad (19)$$

Let us examine the rate of change of the following non-negative function

$$V_k = \tilde{q}_k^\top \Gamma_k^{-1} \tilde{q}_k \quad (20)$$

Denote $\Delta V_k \triangleq V_k - V_{k-1}$. We will first examine the case when $\Gamma_k^{-1} = \Gamma_{k-1}^{-1} + \phi_k \phi_k^\top$. By pre-multiplying $\Gamma_k$ to both sides of this equation, one has:

$$I = \Gamma_k \Gamma_{k-1}^{-1} + \Gamma_k \phi_k \phi_k^\top \quad (21)$$

hence,

$$\Gamma_k \Gamma_{k-1}^{-1} = I - \Gamma_k \phi_k \phi_k^\top \quad (22)$$

Substituting this to (19) yields

$$\tilde{q}_k = \Gamma_k \Gamma_{k-1}^{-1} \tilde{q}_{k-1} \quad (23)$$

Using (23) and (20), we have:

$$\begin{aligned} \Delta V_k &= (\tilde{q}_{k-1}^\top \Gamma_{k-1}^{-1} \Gamma_k) \Gamma_k^{-1} (\Gamma_k \Gamma_{k-1}^{-1} \tilde{q}_{k-1}) \\ &\quad - \tilde{q}_{k-1}^\top \Gamma_{k-1}^{-1} \tilde{q}_{k-1} \\ &= \tilde{q}_{k-1}^\top (\Gamma_{k-1}^{-1} \Gamma_k \Gamma_{k-1}^{-1} - \Gamma_{k-1}^{-1}) \tilde{q}_{k-1} \end{aligned} \quad (24)$$

Since $\Gamma_k = \Gamma_{k-1} - \frac{\Gamma_{k-1} \phi_k \phi_k^\top \Gamma_{k-1}}{1 + \phi_k^\top \Gamma_{k-1} \phi_k}$ in this case, by pre-multiplying then post-multiplying $\Gamma_{k-1}^{-1}$ to both sides, we have:

$$\Gamma_{k-1}^{-1} \Gamma_k \Gamma_{k-1}^{-1} - \Gamma_{k-1}^{-1} = -\frac{\phi_k \phi_k^\top}{1 + \phi_k^\top \Gamma_{k-1} \phi_k} \quad (25)$$

Substituting this to (24) and by using (18), we obtain:

$$\Delta V_k = -\frac{\varepsilon_k^2}{1 + \phi_k^\top \Gamma_{k-1} \phi_k} \quad (26)$$

Thus $\Delta V_k \leq 0$ in this case and the equality occurs when $\varepsilon_k = 0$.

On the other hand, if $\Gamma_k^{-1} = \Gamma_0^{-1}$ then directly substituting this and (19) to $\Delta V_k$ yields:

$$\Delta V_k = -\tilde{q}_{k-1}^\top (\Gamma_{k-1}^{-1} - \Gamma_0^{-1}) \tilde{q}_{k-1}^\top - \varepsilon_k^2 (2 - \phi_k^\top \Gamma_0 \phi_k) \quad (27)$$

Because $\Gamma_{k-1}^{-1} - \Gamma_0^{-1} \geq 0$, $\|\phi_k\| \leq \tau v_M$ and we have assumed $(\tau v_M)^2 \gamma_1 < 2$, it follows that $(2 - \phi_k^\top \Gamma_0 \phi_k) > 0$, thus $\Delta V_k \leq 0$ in this case and the neccessary condition for the equality to occur is $\varepsilon_k = 0$.

Thus, we have established that $V_k > 0$ and $\Delta V_k \leq 0, \forall k > 0$, which means $V_k$ will decrease to some constant value. Therefore $V_k$ is certainly upper bounded by $V_0$. From the definition of $V_k$ and the fact that $\Gamma_k^{-1}$ is bounded by design, $\tilde{q}_k^\top$ and $\hat{q}_k$ are also bounded. This implies that $\varepsilon_k = \tilde{q}_{k-1}^\top \phi_k$ is bounded since $\phi_k$ is bounded. In other words, $\varepsilon_k, \hat{q}_k, \tilde{q}_k \in \ell_\infty$.

To show that $\varepsilon_k \in \ell_2$, we can take the sum of $\Delta V_k$ from $k = 1$ to infinity, notice that $\lim_{k \to \infty} V_k = V_\infty < V_0$, we have:

$$\sum_{k=1}^\infty \Delta V_k = V_\infty - V_0 = -\left( \sum_{k=0}^\infty \varepsilon_k^2 h_k + g_k \right) \quad (28)$$

Notice that $h_k$ and $g_k$ are both non-negative terms that can be expressed explicitly by looking back at (24) and (27). Using this non-negativity we can deduce the following:

$$V_0 - V_\infty = \sum_{k=0}^\infty \varepsilon_k^2 h_k + g_k \geq \sum_{k=0}^\infty \varepsilon_k^2 h_k \quad (29)$$

By the choice of $\tau$, $v_M$ and $\gamma_1$ we can bound $h_k$ from below by **h** defined as:

$$\mathbf{h} \triangleq \min \left\{ 2 - (\tau v_M)^2 \gamma_1, \frac{1}{1 + (\tau v_M)^2 \gamma_1} \right\} \quad (30)$$

Hence (29) can be further developed as:

$$V_0 - V_\infty \geq \sum_{k=0}^\infty \varepsilon_k^2 m_k \geq \sum_{k=0}^\infty \varepsilon_k^2 \mathbf{h} \quad (31)$$

Finally, we can obtain:

$$\sum_{k=0}^\infty \varepsilon_k^2 \leq \frac{V_0 - V_\infty}{\mathbf{h}} < \infty \quad (32)$$

which certifies that $\varepsilon_k \in \ell_2$.

ii. This follows the fact that $V_k > 0$ and $\Delta V_k < 0$, therefore,

$$\lim_{k \to \infty} \Delta V_k = 0 \tag{33}$$

which, due to (26) and (27), implies

$$\lim_{k \to \infty} \varepsilon_k = 0 \tag{34}$$

iii. To show $\|\tilde{q}_k\| \leq \|\tilde{q}_0\|$, notice that since $V_k \leq V_0$, $\forall k > 0$:

$$\tilde{q}_k^\top \Gamma_k^{-1} \tilde{q}_k \leq \tilde{q}_0^\top \Gamma_0^{-1} \tilde{q}_0 \tag{35}$$

Recall that $\Gamma_0^{-1} \leq \Gamma_k^{-1}$, therefore $\tilde{q}_k^\top \Gamma_0^{-1} \tilde{q}_k \leq \tilde{q}_k^\top \Gamma_k^{-1} \tilde{q}_k$, which leads to:

$$\tilde{q}_k^\top \Gamma_0^{-1} \tilde{q}_k \leq \tilde{q}_0^\top \Gamma_0^{-1} \tilde{q}_0 \tag{36}$$

since we have assumed $\Gamma_0 = \gamma_1 I$, $\Gamma_0^{-1} = \gamma_1^{-1} I$ and $\|\tilde{q}_k\| \leq \|\tilde{q}_0\|$ follows directly. ∎

### B. Stability of relative position states

So far we have shown that the estimator states are stable, it remains to show that $p_k$ is also bounded. We are more interested in the boundedness of the relative position between the target and the agent. We denote $\bar{p}_k \triangleq p_k - q$ and seek to prove that this state is bounded. Notice that $p_k - \hat{q}_k = \bar{p}_k - \tilde{q}_k$. We can express (1) and (13) as follows:

$$\bar{p}_{k+1} = \bar{p}_k + \tau s_k u_k, \tag{37}$$

where $s_k$ is defined as:

$$s_k = \begin{cases} 1, & \|u_k\| \leq v_M, \\ v_M/\|u_k\|, & \text{otherwise}, \end{cases} \tag{38}$$

and $u_k$ can be redefined as:

$$u_k = -\beta \bar{p}_k + f(d_k) \sigma_k + \beta \tilde{q}_k. \tag{39}$$

We can now present the following theorem to complete the analysis for stability of the system.

**Theorem 2.** *If $\gamma_1 (\tau v_M)^2 < 2$, $1 - \tau \beta > 0$ and $2 - \tau(\beta - \sigma_M) > 0$ then $\bar{p}_k$ is bounded with the upper bound determined by $\bar{p}_0$, $\hat{q}_0$, $\beta$, $\sigma_M$, $\tau$, $v_M$.*

*Proof.* Define $L_k \triangleq \bar{p}_k^\top \bar{p}_k$ and $\Delta L_k \triangleq L_k - L_{k-1}$. One has:

$$\begin{aligned} \Delta L_{k+1} &= \bar{p}_{k+1}^\top \bar{p}_{k+1} - \bar{p}_k^\top \bar{p}_k \\ &= (\bar{p}_k + \tau \bar{u}_k)^\top (\bar{p}_k + \tau \bar{u}_k) - \bar{p}_k^\top \bar{p}_k \\ &= s_k \tau (s_k \tau u_k^\top u_k + 2 u_k^\top \bar{p}_k) \end{aligned}$$

Notice that since $0 < s_k \leq 1$, we can manipulate the previous inequality as follows:

$$\Delta L_{k+1}/(s_k \tau) \leq \tau u_k^\top u_k + 2 u_k^\top \bar{p}_k \tag{40}$$

Now the right hand side has become much more convenient thanks to the absence of the saturation effect. To make the notation concise hereafter we may omit the time step $k$ and refer to $f(d_k)$ as $f$, $\sigma_k$ as $\sigma$ when it is unambiguously convenient. From (40), one has:

$$\begin{aligned} \Delta L_{k+1}/(s_k \tau) &\leq \\ &\tau(-\beta \bar{p} + f\sigma)^\top (-\beta \bar{p} + f\sigma) + 2\tau \beta (-\beta \bar{p} + f\sigma)^\top \tilde{q} \\ &+ \tau \beta^2 \|\tilde{q}\|^2 + 2(-\beta \bar{p} + f\sigma)^\top \bar{p} + 2\beta \tilde{q}^\top \bar{p} \\ &= \tau \left( \beta^2 \|\bar{p}\|^2 + f^2 \|\sigma\|^2 - 2\beta f \sigma^\top \bar{p} \right) - 2\tau \beta^2 \bar{p}^\top \tilde{q} \\ &+ 2\tau \beta f \sigma^\top \tilde{q} + \tau \beta^2 \|\tilde{q}\|^2 - 2\beta \|\bar{p}\|^2 + 2f\sigma^\top \bar{p} + 2\beta \tilde{q}^\top \bar{p} \\ &\leq (\tau \beta^2 + \tau \sigma_M^2 - 2\beta) \|\bar{p}\|^2 + (2 - 2\tau\beta) f \sigma^\top \bar{p} \\ &+ (2\tau \beta f \sigma^\top \tilde{q} - 2\tau \beta^2 \tilde{q}^\top \bar{p} + 2\beta \tilde{q}^\top \bar{p}) + \tau \beta^2 \|\tilde{q}\|^2 \end{aligned} \tag{41}$$

Notice that we have used the assumption that $f(d_k) < d_k = \|\bar{p}_k\|$ to assert the latest inequality. Let us denote the right hand side of (41) as the sum of the following to terms:

$$\begin{aligned} A &\triangleq (\tau \beta^2 + \tau \sigma_M^2 - 2\beta) \|\bar{p}\|^2 + (2 - 2\tau\beta) f \sigma^\top \bar{p} \\ B &\triangleq (2\tau \beta f \sigma^\top \tilde{q} - 2\tau \beta^2 \tilde{q}^\top \bar{p} + 2\beta \tilde{q}^\top \bar{p}) + \tau \beta^2 \|\tilde{q}\|^2 \end{aligned} \tag{42}$$

Using the the assumption that $(1 - \tau\beta) > 0$, we have $(2 - 2\tau\beta) f \sigma^\top \bar{p} \leq (2 - 2\tau\beta) \|\bar{p}\| \sigma^\top \bar{p}$. Then since $\sigma^\top \bar{p} \leq \sigma_M \|\bar{p}\|$, we can derive the following chain of inequalities:

$$\begin{aligned} A &\leq \left( \tau \sigma_M^2 + \tau \beta^2 - 2\beta + 2\sigma_M - 2\tau \beta \sigma_M \right) \|\bar{p}\|^2 \\ &\leq \left( \tau (\beta - \sigma_M)^2 - 2(\beta - \sigma_M) \right) \|\bar{p}\|^2 \\ &\leq (\beta - \sigma_M)(\tau(\beta - \sigma_M) - 2) \|\bar{p}\|^2 \end{aligned} \tag{43}$$

Now we invoke the assumptions that $2 - \tau(\beta - \sigma_M) > 0$ to arrive at the following:

$$A \leq -(\beta - \sigma_M)(2 - \tau(\beta + \sigma_M)) \|\bar{p}\|^2 \tag{44}$$

From Theorem 1 we have $\tilde{q}_k \in \ell_\infty$, thus $\exists M_1, \|\tilde{q}_k\| \leq M_1$, $\forall k$. Thus, one can assert the following inequality:

$$B \leq (2\tau \beta \sigma_M M_1 + 2\tau \beta^2 M_1 + 2\beta M_1) \|\bar{p}\| + \tau \beta^2 M_1^2 \tag{45}$$

Thus from (41), (44), (45) we can obtain the following:

$$\Delta L_{k+1} \leq s_k \tau \left( -a \|\bar{p}\|^2 + b \|\bar{p}\| + c \right) \tag{46}$$

where

$$\begin{aligned} a &\triangleq -(\beta - \sigma_M)(2 - \tau(\beta + \sigma_M)) > 0 \\ b &\triangleq (2\tau \beta \sigma_M + 2\tau \beta^2 + 2\beta) M_1 \geq 0 \\ c &\triangleq \tau \beta^2 M_1^2 \geq 0 \end{aligned} \tag{47}$$

Since inside the bracket on right hand side of (46) is a standard quadratic function with a negative leading coefficient, we can conclude that

$$\exists M_2, \Delta L_{k+1} \leq 0, \forall \|\bar{p}_k\| \geq M_2 \tag{48}$$

Clearly $M_2$ depends on the coefficients $a$, $b$, $c$, which depend on $\beta$, $\sigma_M$, $\tau$ and $\|\tilde{q}_0\|$. Thus we can assert the upper bound of $\|\bar{p}_k\|$ as follows:

$$M = \max\{M_2 + \tau v_M, \|\bar{p}_0\|\} \tag{49}$$

Notice that $M_2 + \tau v_M$ instead of $M_2$ is considered as the upper bound because if $\|\bar{p}_0\| < M_2$, then $\Delta L_{k+1}$ can still be positive and hence $\|\bar{p}_k\|$ may still increase up to $M_2 + \tau v_M$ before $\Delta L_{k+1}$ becomes non-positive. On the other hand if $\|\bar{p}_0\| \geq M_2 + \tau v M$ then we are certain $\Delta L_{k+1}$ is non-positive and $\|\bar{p}_k\| \leq \|\bar{p}_0\|$, $\forall k > 0$. ∎

**Remark 1.** As we have assumed that $1 - \tau\beta > 0$, it suffices to have $\beta - \sigma_M$ to guarantee $2 - \tau(\beta - \sigma_M) > 0$. This is because $2 - \tau(\beta - \sigma_M) = (1 - \tau\beta) + 1 + \sigma_M > 0$.

## IV. SIMULATION

To fully verify the capability of the proposed algorithm in this section we provide a simulation in a 3D scenario. The parameters for the simulation are listed in Table I and satisfy all of the requirements to guarantee stability and convergence as pointed out by the aforementioned Theorems. It should be noticed that the time step is set at $1s$, which is reasonable for a system to have the dynamics approximated by a single integrator. The initial value of the process $\sigma_k$ is chosen as $\sigma_0 = \begin{bmatrix} r_1, & 0, & r_2\cos(0.1), & r_2\sin(0.1) \end{bmatrix}^\top$ and $r_1 = r_2 = 0.25$ as listed in Table I.

TABLE I
PARAMETERS USED IN THE SIMULATION.

| Dynamics | Control | $\sigma_k$ | Estimation |
|---|---|---|---|
| $\tau = 1$ | $\lambda = 0.1$ | $r_1 = r_2 = 0.25$ | $\Gamma = I$ |
| $v_M = 0.5$ | $\beta = 0.5$ | $\omega = 2\pi/18$ | |

### A. Static target

As can be seen in Fig. 2, the target is at the location $[-5, 2, 10]$ and the agent's initial position is at $[-35, -35, 1]$, which is also used as the initial estimate $\hat{q}_0$. The entire trajectory is plotted in 3D in Fig. 4. It can be seen that as time progresses, all of the system states converge to $q$. In Fig. 3 we can see that the distance between the agent and the target approaches zero as time moves on. As Theorem 1 has pointed out, we can see in Fig. 5 that the function $V_k$ is positive definite, $\Delta V_k$ is negative semi-definite and $V_k$ decreases to zero as time approaches infinity.

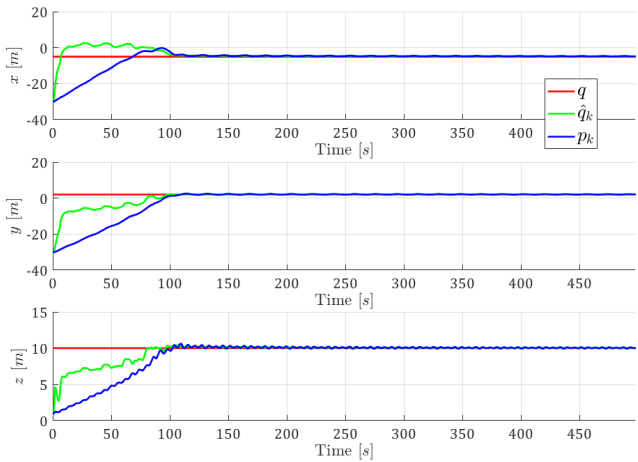

Fig. 2. Time evolution of system's states.

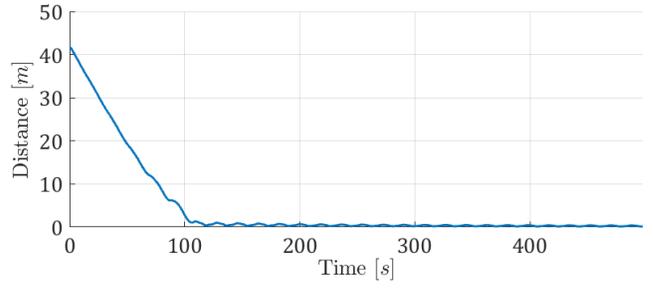

Fig. 3. Time evolution of the distance to target.

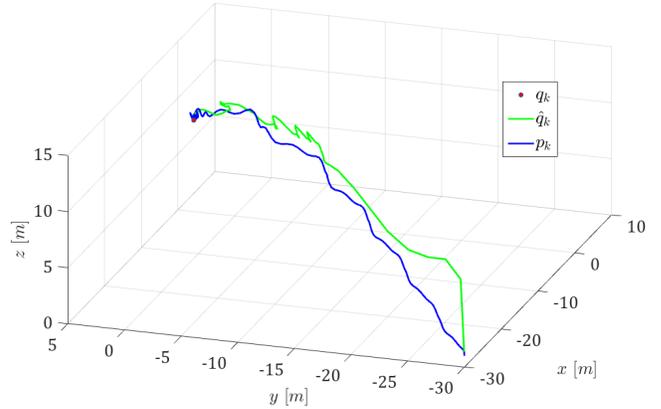

Fig. 4. Trajectory of the agent and estimate of the target location.

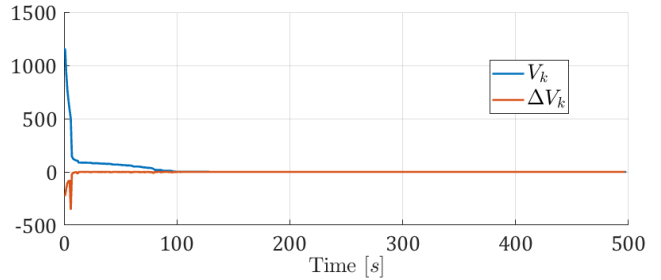

Fig. 5. Time evolution of the function $V_k$ and $\Delta V_k$.

### B. Moving target

In this section we intentionally violate the assumption that the target's location is static to examine the robustness of the adaptive control scheme. For this purpose we use the following trajectory for the target

$$q_k = q_0 + \begin{bmatrix} 10\cos(\omega k + 0.1) \\ 10\sin(\omega k + 0.1) \\ 2.5\cos(3\omega k + 2) \end{bmatrix} + k\tau \begin{bmatrix} -0.05 \\ 0.05 \\ 0.025 \end{bmatrix}. \quad (50)$$

where $\omega = 6.136 \times 10^{-3}$.

Fig. 6 shows the trajectories of the system states with the same initial position for the agent and the estimate of the target location. As time progresses, the target's position drifts as dictated by (50). It can be seen that the agent can maintain a bounded distance to the target in Fig. 3 and the estimator state $\hat{q}_k$ can also track the target's true position, as can be seen in Fig. 4.

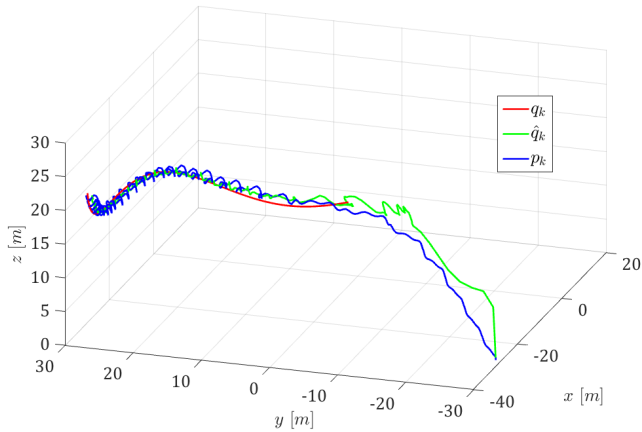

Fig. 8. Trajectory of the agent and estimate of the target location with a moving target.

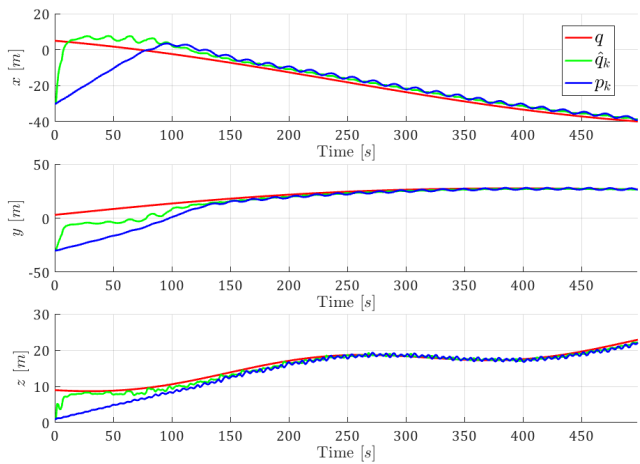

Fig. 6. Time evolution of system's states with moving target.

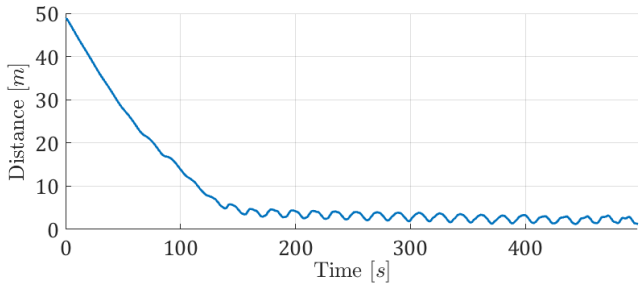

Fig. 7. Time evolution of the distance to target with moving target.

## V. CONCLUSION

In this paper we have studied the problem of target tracking using the least-square based optimization technique. Analysis on the closed-loop system was carried out to stipulate important parametric conditions to achieve stability of all system states. Via simulation, convergence of the agent's position can be observed regardless of initial conditions.

For future works, we look forward to establishing the asymptotic convergence of the agent-target relative position.

It would be desirable to consider the problem where the final location to converge to is distinct from the location of the landmark. Cooperative localization of multiple agents using distance measurements to the landmark and between each other would also be an interesting problem.

ACKNOWLEDGEMENT

We would like to thank Dr. Li Xiuxian for his insightful comments and discussions that helped materialized the results in this paper.